\newcommand{\be}{\begin{equation}}
\newcommand{\ee}{\end{equation}}
\newcommand{\bea}{\begin{eqnarray}}
\newcommand{\eea}{\end{eqnarray}}
\newcommand{\ben}{\begin{eqnarray*}}
\newcommand{\een}{\end{eqnarray*}}

\newcommand{\ii}{\mathop{\mathrm i}\nolimits}

\documentclass{elsart5}

 \usepackage{graphicx}

\usepackage{amssymb}
\usepackage{amsmath}

\begin{document}

\begin{frontmatter}

\title{Ergodicity of the extended anisotropic 1D Heisenberg 
model:\\ response at low temperatures}

\author{Evgeny Plekhanov\corauthref{cor1}}
\ead{plekhanoff@sa.infn.it}
\ead[url]{http://scs.sa.infn.it/plekhanoff/}
\corauth[cor1]{}
\author{Adolfo Avella}
\author{Ferdinando Mancini}
\address{Dipartimento di Fisica ``E.R. Caianiello'' - Unit\`{a}
CNISM di Salerno \\
Universit\`{a} degli Studi di Salerno, I-84081 Baronissi (SA),
Italy}


\begin{abstract}
We present the results of exact diagonalization calculations
of the isolated and isothermal on-site static susceptibilities 
in the anisotropic extended Heisenberg model on a linear chain
with periodic boundary conditions.
Based on the ergodicity considerations we conclude
that the isothermal susceptibility will diverge as $T\to 0$
both in finite clusters and in the bulk system in two
non-ergodic regions of the phase diagram of the system.
\end{abstract}

\begin{keyword}
\PACS 75.40.C\sep 71.27\sep 75.10.D
\KEY  Ergodicity\sep Magnetic response\sep One-dimensional spin systems
\end{keyword}

\end{frontmatter}

It is often thought that the ergodicity - the property
to explore the whole Hilbert space during time
evolution - is a natural attribute of a physical system
and, moreover, that its possible violation is hard to observe.
Ergodicity assumptions
are usually made when passing through various self-consistent
schemes in determining susceptibilities, correlation functions {\sl etc}.
Nevertheless, non-ergodic behavior may manifest under certain 
circumstances. The purpose of this article is to show
how non ergodicity can significantly change
the properties of the system and may become in principle observable.

When dealing with the static response of a quantum system two
definitions of the susceptibility are usually considered:
the static "isolated", or "Kubo susceptibility", $\chi_0$ and 
the isothermal one $\chi^T$~\cite{Kubo}. These two definitions correspond
to different environmental conditions under which
the response is measured. Namely, $\chi_0$ is derived upon
the assumption that the system is isolated, while in order
to measure $\chi^T$ one has to maintain the system in thermal
equilibrium with an external bath at a given temperature.
As it was pointed out by Kubo and Suzuki~\cite{Suzuki},
a necessary and sufficient condition for these two response
functions to coincide is the ergodicity of the system.
By considering as a specific example the response function
of an operator $A$ to a perturbation coupled to $A$ itself,
the isolated Kubo susceptibility will be:
\be
   \chi_0 = -\ii
   \int_{0}^{\infty} dt
   \langle \left[
   A(t), A(0)
   \right]
   \rangle
   \label{aaa}
\ee
while the isothermal one:
\be
   \chi^T = \int_0^{\beta} d\lambda 
   \langle A(-\ii\lambda) A \rangle
   -\beta \langle A \rangle^2.
\ee
Their difference, as shown in~\cite{Kubo,Suzuki},
is given by:
\be
   \chi^T - \chi_0 = \beta ( \Gamma - \Gamma^{erg}),
   \label{suz_th}
\ee
where we have introduced the quantity $\Gamma$:
\be
   \Gamma = 
   \lim_{t\to\infty}
   \langle
   A(t) A(0)
   \rangle
\ee
and its ergodic value
\be
   \Gamma^{erg} = \langle A \rangle^2.
\ee
When $\Gamma\neq \Gamma^{erg}$ the response
is said to be non-ergodic and the two susceptibilities
differ. In this case, non-ergodicity becomes 
observable as it provides a diverging contribution
to $\chi^T$ in the limit $T\to 0$.

\begin{figure*}[ht]
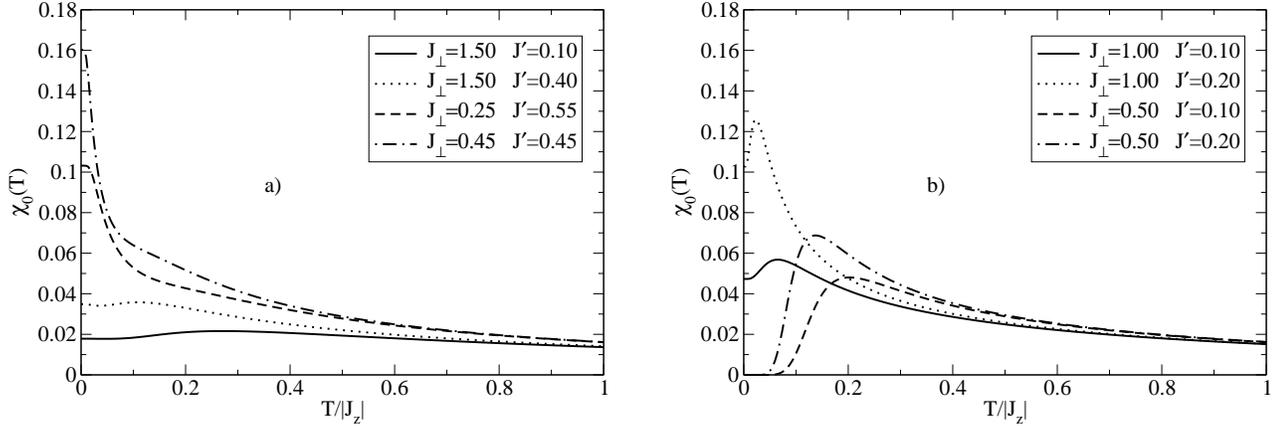

\includegraphics[width=8.0cm]{chi_ii_erg.eps}
\hspace{0.6cm}
\includegraphics[width=8.0cm]{chi_ii_non_erg.eps}
\caption{
On-site magnetic susceptibility in the various regions
from the phase diagram (see~\cite{ours}) as a function
of temperature for the Hamiltonian~(\ref{ham}) on 
a linear chain of 14 sites.
Panel a) collects the results for the points from 
Ergodic I (solid and dotted lines) and 
Ergodic II phases (dashed and dashed-dotted lines) 
while the panel b) those from Non-Ergodic I 
(solid and dotted lines) and the Non-Ergodic II phases
(dashed and dashed-dotted lines).
}
\label{fig1}
\end{figure*}

In this paper we show the results for the isolated susceptibility
in the extended anisotropic Heisenberg model on a linear chain 
of $14$ sites with periodic boundary conditions. 
The susceptibility has been calculated using the spectrum of the
system obtained by means of the exact numerical diagonalization.
The Hamiltonian of the system reads as:
\begin{multline} \label{ham}
   H = -J_z \sum_i S^z_i S^z_{i+1} \\
   +J_\perp \sum_i ( S^x_i S^x_{i+1} + S^y_i S^y_{i+1}) +J^\prime
   \sum_i \mathbf{S}_i \mathbf{S}_{i+2}.
\end{multline}
We fix $J_z>0$ so that the $z$-component of the nearest-neighbors
exchange is ferromagnetic and use $J_z$ as the energy unit.
We use $S^z_{i}$ as the operator $A$ in~(\ref{aaa}): 
\be
   \chi_0(i,i) = -\ii
   \int_{0}^{\infty} dt
   \langle
   \left[
   S^z_i(t), S^z_i(0)
   \right]
   \rangle.
\ee
Because of the translational invariance $\chi_0(i,i)$ is
independent of the site $i$.

Previously~\cite{Bak},\cite{ours} we have already
studied the question of ergodicity of the response of $S^z_i$
in the Hamiltonian~(\ref{ham}). 
We have constructed a phase diagram of~(\ref{ham})
in the plane $J^\prime-J_\perp$ for both zero and finite temperatures
using as large as 26 sites clusters.
We have concluded that at $T=0$
and in the bulk limit there are two non-ergodic regions and 
two ergodic ones (see Fig.1 in~\cite{ours}). 
A transition zone exists at finite sizes, which probably
would become a transition line in the bulk system. At $T>0$ the
finite-size scaling indicated that the system is always ergodic.

On Fig.~\ref{fig1} the Kubo susceptibility is plotted as a function 
of temperature for the ergodic (panel a) and non-ergodic (panel b)
regions from the phase diagram obtained in~\cite{ours}. 
While at high temperatures $\chi_0(T)\sim 1/T$,
as it should be after the Curie law, the most interesting behavior is
concentrated at low temperatures. One can easily show that when $T\to 0$
the Kubo susceptibility can be rewritten as:
\be
   \chi_0(T) = \frac{2}{N}
   \sum_{l=1}^{N}
   \sum_{
      E_n > E_0
   }
   \frac{
   |\langle
   0, l | S^z_i | n
   \rangle|^2}
   {E_n - E_0} + \Omega(T),
   \label{2-order}
\ee
where $N$ is the number of
degenerate ground states ($|0, l\rangle$, $l=1,\ldots N$),
$|n\rangle$ is the $n-$th excited eigenstate of~(\ref{ham})
and $\Omega(T)$ goes exponentially to zero when $T\to 0$.
(\ref{2-order}) is nothing else but the result of the 
perturbation theory up to the second order.
One can see from~(\ref{2-order}) that $\chi_0(0)$ is 
non-singular and in general non-zero since all the terms under the
sum have the same sign.
This is not the case in the phase Non-ergodic I (see the 
dashed and dashed-dotted lines on the Fig.~\ref{fig1}b)) 
since in this phase the ground state
is doubly degenerate with all spins either up or down. Such states are
eigenstates of $S^z_i$ and therefore the matrix elements 
in~(\ref{2-order}) vanish. All the other phases have their ground
states connected to the rest of the Hilbert space by $S^z_i$ 
so that for them $\chi_0(0)\neq 0$.

From our data for the isolated susceptibility and by
using~(\ref{suz_th}), we can find the
temperature dependence of $\chi^T$ as well.
It is clear from~(\ref{suz_th}) that in both ergodic
phases (Ergodic I and Ergodic II) $\chi^T(T)=\chi_0(T)$.
On the contrary, the isothermal susceptibility will
diverge
at low $T$ as $\Gamma/T$, where $\Gamma=1/12+1/6L$ for 
the phase Ergodic I on a cluster of $L$ sites, 
and $\Gamma=1/4$ for the phase Ergodic II, independently on $L$.
Based on the finite-size scaling for $\Gamma$ 
made in~\cite{ours} these results will hold also in the thermodynamic
limit.

In conclusion, we have studied the on-site static magnetic response
of~(\ref{ham}), considering both definitions of the susceptibility.
While for the isolated one the non ergodicity enters implicitly
through the degeneracy of the eigenvalues, the isothermal
susceptibility contains an explicit term diverging at low-$T$ in the
case of non-ergodic phases. In real materials, however, this
divergence might not be observable due to perturbations such as
dipole-dipole interactions which destroy the conservation of
$z$-component of the total spin and ensure the ergodicity.

\end{document}